\documentclass[aps,preprint]{revtex4}
\usepackage{amssymb}
\usepackage{amsfonts}
\usepackage{amsmath}
\usepackage{graphicx}

\begin{document}
\title[Demonstration of Levinson theorem]{Reference potential approach to the inverse Schr\"{o}dinger problem: explicit
demonstration of Levinson theorem and a solution scheme for Krein equation}
\author{Matti Selg}
\affiliation{Institute of Physics of the University of Tartu, Riia 142, 51014 Tartu, Estonia}

\begin{abstract}
A recently proposed reference potential approach to the inverse
Schr\"{o}dinger problem is further developed. As previously, theoretical
developments are demonstrated on example of diatomic xenon molecule in its
ground electronic state. An exactly solvable reference potential for this
quantum system is used, which enables to solve the related energy eigenvalue
problem exactly. Moreover, the full energy dependence of the phase shift can
also be calculated analytically, and as a particular result, full agreement
with Levinson theorem has been achieved and explicitly demonstrated. In
principle, this important spectral information can be reused to calculate an
improved potential for the system, and such possibilities are discussed in the
paper. Aiming at this goal, one may calculate an auxiliary potential with no
bound states, whose spectral density for positive energies is exactly the same
as that of the reference potential. To this end, one may solve Krein equation,
which in the present context is simpler than using Gelfand-Levitan method.
General solution of Krein equation can be expressed as a Neumann series.
Convergence of this series of multi-dimensional integrals at distances not
close to the origin is hard to achieve without a simple asymptotic formula for
calculating the kernel of Krein equation. As proven in this paper, such an
asymptotic formula exists, and its parameters can be easily ascertained.

\end{abstract}
\maketitle

\section{Introduction}

Celebrated Levinson theorem \cite{Levinson1,Levinson2}%
\begin{equation}
\delta_{l}(0)-\delta_{l}(\infty)=n_{l}\pi,
\end{equation}
originally associated with spherically symmetric potentials in three
dimensions, is simple in form, but fundamental in content. Indeed, Eq. (1)
reveals a correlation between the main characteristics of the continuous and
discrete energy spectra, the phase shift $\delta_{l}(E)$ ($E>0$) and the total
number ($n$) of bound states ($E<0$). The quantity $n_{l}=n+1/2,$ if the
angular momentum $l=0$ and a so-called half-bound state exists at $E=0,$ and
$n_{l}=n$ in any other case. As can be shown, the solution of the
Schr\"{o}dinger equation for $l=0$ and $E=0$ (if one exists) is not
normalizable, but it still gives contribution to Levinson theorem, a half of
that compared with any "normal" bound state. Therefore, the term "half-bound
state" introduced by Newton \cite{Newton} is fitting and illustrative for this
rather special solution.

A lot of proofs of the Levinson theorem have been given by several authors
using different methods (see, e.g., \cite{DongHouMa} for a brief overview),
and various generalizations of Eq. (1) are known. For example, Levinson
theorem has been adjusted to noncentral \cite{Newton} and nonlocal
\cite{Martin} potentials (including bound states with $E>0$), electron-atom
scattering \cite{Rosenberg}, relativistic equations in one \cite{Lin,Dong},
two \cite{DongHouMa} and more \cite{Jiang} dimensions, etc. In spite of the
vast literature on the subject, one hardly can find any examples of
calculating the full energy dependence (from 0 to $\infty$) of the phase
shift, which would explicitly demonstrate the Levinson theorem. Such a
computation for a particular quantum system is a challenge even nowadays, but
as we demonstrate below, the computational-technical difficulties can be
overcome, if one is able to construct a reasonable reference potential for the
system. Moreover, for any $E>0,$ the phase shift can be ascertained exactly,
if one uses an exactly solvable reference potential. In the present context
"exact solubility" means that the problem can be solved with the help of
solely analytic tools, without applying methods of numerical integration.

One-dimensional inverse theory (including Levinson theorem) can be applied to
diatomic molecules, since a two-particle problem can be always reduced to a
one-particle problem in an effective spherically symmetric field where the
angular momentum $l$ is replaced by the rotational quantum number $J$. On this
basis, the reference potential method developed in this paper is illustrated
on example of diatomic xenon molecule (Xe$_{2}$) in its ground electronic
state. To avoid excessive technical difficulties and concentrate on the
essence of the approach, only the rotationless case ($J=0$) is analyzed.

Complete knowledge of the phase shift is one of the three obligatory
preconditions for the unique solution of the inverse Schr\"{o}dinger problem
for a quantum system. In addition, one has to ascertain full energy spectrum
of the bound states $E_{m}<0$ $(m=1,...,n)$ and $n$ associated real parameters
$C_{m}$ that uniquely fix their normalization \cite{Chadan}. In principle, all
these data can be obtained experimentally, but unfortunately, this is almost
unachievable for a real system. In this connection, one may ask: is it still
possible to get some practical benefit from the rigorous methods of the
inverse quantum theory? One option, as already mentioned, is to construct a
reference potential based on the known data on the system's properties. Since
the reference potential is fixed, all its spectral parameters are determined
\textit{a priori}. Naturally, there is no need to regain a potential which is
already known by definition, but the idea is to use the calculated reference
parameters as zeroth approximations to the real characteristics. Using these
approximate spectral characteristics, it is possible, at least in principle,
to build up a more realistic potential for the system. For example, one can
construct a Bargmann potential \cite{Bargmann1, Bargmann2} whose bound states
coincide with the actually observed discrete energy levels, or adjust the
potential more adequately to the available scattering data. These
possibilities will be discussed in this paper, which is organized as follows.

In Section II we first specify the analytic form of the reference potential,
and then describe the details of calculating the phase shift for the full
range of scattering states. Having ascertained the phase shift, one can
calculate another important spectral characteristic, Jost function, and then
fix spectral density (the terms to be specified below) for the reference
potential. In Section III we demonstrate how these spectral characteristics
can be reused to calculate a special auxiliary potential with no bound states,
applying Krein method of solving the inverse scattering problem. Starting from
this auxiliary potential, one can calculate an improved (Bargmann) potential
for the system, as already mentioned. Several possibilities of this kind are
discussed in Section IV. Finally, the main results of the work are summarized
in Section V.

\section{Full energy dependence of the phase shift}

According to the scheme outlined in the previous section, our first goal is to
construct an exactly solvable (in the above specified sense) reference
potential for the system. A good choice for this purpose, as explained in
detail elsewhere \cite{Selg1, Selg2}, is a multi-component potential composed
of several smoothly joined Morse type pieces%
\begin{equation}
V(r)=V_{k}+D_{k}\left[  \exp(-\alpha_{k}(r-r_{k}))-1\right]  ^{2},\text{ }%
r\in(0,\infty),
\end{equation}
where $V_{k},$ $D_{k},$ $\alpha_{k}$ and $r_{k}$ are some constants (not
definitely positive), and the subscript $k$ corresponds to different
components smoothly joined at some suitably chosen boundary points $X_{k+1}$.
A five-component reference potential for Xe$_{2}$ has been constructed
\cite{Selg2}, which is in good agreement with all known spectroscopic data. In
this paper the emphasis is put on principles, not on specific details.
Therefore, a more simple (but still reasonable) reference potential of three
Morse components ($k=0,1,2$) is used. The most internal of them ($k=0$)
represents a so-called pseudo-Morse potential. It means that the tiny
potential well corresponding to this component (if taken separately) is just
of the limit depth to completely lose the discrete energy spectrum.
Consequently, $D_{0}=\hbar^{2}\alpha_{0}^{2}/(8m)$ ($m$ being the reduced mass
of the pair of atoms), so that only three independent parameters remain. The
central component ($k=1$) is an ordinary Morse potential, while the most
external one ($k=2$) is a "reversed" Morse potential with the parameter
$D_{2}$ being negative. By introducing a "reversed" component one artificially
creates a potential barrier in the long-distance range. This might seem
unphysical and unjustified, but the point is that the height of this
artificial hump approaches zero as the parameter $r_{2}$\ approaches infinity.
Therefore, taking a sufficiently large $r_{2}$, the hump becomes almost
insignificant, while the analytic treatment remains simple and flexible.
Almost equivalently, one may fix $V(r)=0$, if $r\geq r_{2}$. The parameters of
the reference potential are given elsewhere \cite{SelgArXiv1}, and the curve
itself is shown in Fig. 1. The described construction based on Morse type
functions is not occasional. On the contrary, this is one of the very few
options to easily get two linearly independent analytic solutions of the
Schr\"{o}dinger equation for the whole physical domain.

As already mentioned, the phase shift for the specified reference potential
for any $E\in(0,\infty)$ can be ascertained analytically. To demonstrate this,
let us briefly recall the solution scheme for Morse-type potentials (see
\cite{Selg1,SelgArXiv1} for more details). First one introduces dimensionless
variables $y_{k}=2a_{k}\exp(-\alpha_{k}(r-r_{k})),$ where$\ $the subscripts
$k=0,1,2$ have the same meaning as specified above, and $a_{k}=\sqrt{2mD_{k}%
}/(\hslash\alpha_{k})$. The relevant Schr\"{o}dinger equations then read
\begin{equation}
\frac{d^{2}\Psi(y_{k})}{dy_{k}^{2}}+\frac{1}{y_{k}}\frac{d\Psi(y_{k})}{dy_{k}%
}+\left[  -\frac{\mu_{k}^{2}}{y_{k}^{2}}\pm\left(  \frac{a_{k}}{y_{k}}%
-\frac{1}{4}\right)  \right]  \Psi(y_{k})=0,
\end{equation}
where the plus sign in square brackets corresponds to the subscripts $k=0$ and
$k=1$, and minus -- to $k=2$. The quantities $\mu_{k}^{2}$ are defined as
follows: $\mu_{k}^{2}=(a_{k}^{2}/D_{k})\cdot(V_{k}+D_{k}-E).$ Note that
$D_{2}<0$ and $a_{0}=1/2$. Next, Eq. (3) is converted to confluent
hypergeometric equation, whose two linearly independent solutions can be
always ascertained analytically. For example, general solution for the
internal region $r\leq X_{1}$ can be written as $\Psi_{0}(r)=N_{1}\Psi
_{0}^{(1)}(r)+N_{2}\Psi_{0}^{(2)}(r)$, where $\Psi_{0}^{(1)}(r)=A_{0}%
(y_{0})\cos\left[  B_{0}(y_{0})-\varphi_{0}-\alpha_{0}\mu_{0}r\right]  $,
$\Psi_{0}^{(2)}(r)=A_{0}(y_{0})\sin\left[  B_{0}(y_{0})-\varphi_{0}-\alpha
_{0}\mu_{0}r\right]  $, the functions $A_{0}(y_{0})$ and $B_{0}(y_{0})$ are
defined as follows:%
\begin{equation}
A_{0}(y_{0})e^{iB_{0}(y_{0})}\equiv\exp(-y_{0}/2)_{1}F_{1}(i\mu_{0};2i\mu
_{0}+1;y_{0}),
\end{equation}
and $_{1}F_{1}(a;c;x)$ denotes Kummer confluent hypergeometric function. As
demonstrated elsewhere \cite{SelgArXiv1}, special solution $\Psi_{0}^{(2)}(r)$
is almost insignificant if a classically forbidden region exists, i.e., if
$E<V(0)$ (note that $V(0)$ is very large, but still finite in the framework of
our approach). Correspondingly, the correct physical solution (apart from
normalization) practically reduces to the particular solution $\Psi_{0}%
^{(1)}(r).$Therefore, the phase parameter (NB! not yet the actual phase shift)
$\varphi_{0}\equiv$ $\alpha_{0}\mu_{0}r_{0}-\arg\left[  \dfrac{\Gamma
(2i\mu_{0})}{\Gamma(i\mu_{0})}\right]  $ for the region $E\lesssim V(0)$ can
be calculated very easily \cite{Selg1,Selg2}:%
\begin{equation}
\varphi_{0}=\mu_{0}\left(  \alpha_{0}r_{0}+1-\ln2-\frac{1}{2}\ln(1+4\mu
_{0}^{2})\right)  +\frac{1}{2}\int\limits_{0}^{\infty}\left(  \coth t-\frac
{1}{t}\right)  e^{-t}\sin(2\mu_{0}t)\frac{dt}{t},
\end{equation}
where the integral can be conveniently evaluated \cite{Selg1}%
\begin{equation}
I\equiv\int\limits_{0}^{\infty}\left(  \coth t-\frac{1}{t}\right)  e^{-t}%
\sin(2\mu_{0}t)\frac{dt}{t}=\int\limits_{0}^{T}e^{-t}\sin(\pi\frac{t}%
{T})f(t)dt,\text{ }T=\frac{\pi}{2\mu_{0}},
\end{equation}%
\[
f(t)=\frac{\coth t-\frac{1}{t}}{t}-e^{-T}\frac{\coth(t+T)-\frac{1}{t+T}}%
{t+T}+e^{-2T}\frac{\coth(t+2T)-\frac{1}{t+2T}}{t+2T}-...
\]
Another equivalent formula for this quantity reads \cite{Selg2}%

\begin{equation}
I=\sum_{n=1}^{\infty}I_{n},\text{ \ }I_{n}=\frac{(-1)^{n-1}2^{2n}B_{n}%
}{(2n)(2n-1)(1+4\mu_{0}^{2})^{2n-1}}\sum_{k=0}^{n-1}(-1)^{k}\binom{2n-1}%
{2k+1}(2\mu_{0})^{2k+1}.
\end{equation}
Here $B_{n}$ denotes the $n$-th order Bernoulli number. Since two different
definitions of the Bernoulli numbers are widely used \cite{MathWorld}\ (and
sometimes confused), we have to specify that the "old-style" definition is
meant here, i.e. $B_{1}=1/6,$ $B_{2}=1/30,$ $B_{3}=1/42,$ etc.

The simple scheme just described cannot be used in the region $E\gtrsim V(0)$
where special solutions $\Psi_{0}^{(1)}(r)$ and $\Psi_{0}^{(2)}(r)$ become
equally important. Nevertheless, the analytic treatment can be preserved.
Indeed, phase shift is related to regular solutions of the Schr\"{o}dinger
equation, which means that the physically correct linear combination
of$\ \Psi_{0}^{(1)}(r)$ and $\Psi_{0}^{(2)}(r)$ should vanish as
$r\rightarrow0.$ From this one immediately concludes that the correct wave
function $\Psi(r)\sim\sin\{\alpha_{0}\mu_{0}r+B_{0}\left[  y_{0}(0)\right]
-B_{0}\left[  y_{0}(r)\right]  \}.$ According to Eq. (4) it means that one
first has to ascertain the argument of the complex function $_{1}F_{1}\left[
i\mu_{0};2i\mu_{0}+1;y_{0}(r)\right]  $ at $r=0,$ to calculate the phase shift
for the region $E\gtrsim V(0).$ This is a really complicated task, if one
tries to use primary definition%
\begin{equation}
_{1}F_{1}(a;c;z)\equiv1+\dfrac{a}{c}\dfrac{z}{1!}+\dfrac{a(a+1)}{c(c+1)}%
\dfrac{z^{2}}{2!}+....,
\end{equation}
because $z\equiv y_{0}(0)$ is a very large quantity, and as a result, the
series converges very-very slowly. It means that extremely high precision of
computations is needed to get the correct result. The convergence can be
essentially improved due to the relationship between the arguments, $c=2a+1$,
specific to the pseudo-Morse approximation \cite{Selg2}, namely%
\begin{equation}
A_{0}(z)e^{iB_{0}(z)}=\sum_{n=0}^{\infty}\frac{\left(  z/4\right)  ^{2n}%
}{n!\left(  i\mu_{0}+1/2\right)  _{n}}\left(  1-\frac{z/4}{i\mu_{0}%
+1/2+n}\right)  ,
\end{equation}
where $(a)_{n}\equiv\Gamma(a+n)/\Gamma(a)$ is the Pochhammer symbol. There
exists even third, and in the present context, most appropriate alternative to
calculate the needed quantity $B_{0}(z).$ Namely, $_{1}F_{1}(a;c;z)$ can be
expressed in terms of Buchholz polynomials \cite{Buchholz} $p_{n}(b,z)$, and
as has been shown \cite{Abad}, these polynomials in turn can be given as sums
of products of polynomials in $b$ (related to $E$) and $z$ (related to $r$),
separately. In the present case this leads to the formula%
\begin{equation}
A_{0}(z)e^{iB_{0}(z)}=\sum_{n=0}^{\infty}p_{n}(2i\mu_{0}+1,z)\frac{_{0}%
F_{1}(;a;-z/2)}{2^{n}\left(  2i\mu_{0}+1\right)  _{n}},
\end{equation}
where $_{0}F_{1}(;a;z)\equiv\sum\limits_{n=0}^{\infty}\dfrac{z^{n}}{n!(a)_{n}%
},$ and the Buchholz polynomials read%
\begin{equation}
p_{n}(b,z)=\frac{(iz)^{n}}{n!}\sum_{s=0}^{[n/2]}\binom{n}{2s}f_{s}%
(b)g_{n-2s}(z),
\end{equation}
with $f_{0}(b)=1,$ $f_{s}(b)=-(b/2-1)\sum\limits_{r=0}^{s-1}\binom{2s-1}%
{2r}\dfrac{4^{s-r}B_{s-r}}{s-r}f_{r}(b),$ and $g_{0}(z)=1,$ $g_{m}%
(z)=-\dfrac{iz}{4}\sum\limits_{k=0}^{[\frac{m-1}{2}]}\binom{m-1}{2k}%
\dfrac{4^{k+1}B_{k+1}}{k+1}g_{m-2k-1}(z)$ ($s=1,2,...$, and $m=1,2,...$).
Again, as above, $B_{n}$ denotes the $n$-th order "old-style" Bernoulli number.

Thus we have described all details of calculating the phase parameter within
the pseudo-Morse approximation, which is used for $r\in\left[  0,X_{1}\right]
.$ Having fixed this parameter, one can easily calculate the real phase shift,
using the general asymptotic formula $\Psi(r)\sim\sin\left[  kr+\delta
(k)\right]  $ ($k\equiv\sqrt{2mE}/\hbar$), and the continuity conditions for
the wave function and its derivative at the boundary points $X_{k+1}$. The
whole procedure is based on exact analytic formulae, and there is no need for
numerical solution of the well-known phase equation \cite{Morse}%
\begin{equation}
\delta^{\prime}(r,k)=-\dfrac{\sqrt{2m}V(r)}{\hbar k}\sin^{2}\left[
kr+\delta(r,k)\right]  ,\text{ }\delta(0,k)=0,
\end{equation}
which would be a much more difficult task, since the phase shift can only be
obtained as the limit $\delta(k)=\lim\limits_{r\rightarrow\infty}\delta(r,k)$.
We still have to mention that it is not so easy to make a reasonable use of
Eqs. (9) or (10), in order to ascertain the quantity $B_{0}\left[
y_{0}(0)\right]  .$ Although convergence of these complex series is much
better compared with Eq. (8), one definitely has to overcome the barrier of
standard double (or fourfold) precision of high level programming languages,
because much higher accuracy is actually needed. Fortunately, this is not any
serious problem nowadays. Fast and powerful computational tools are freely
available, such as high-precision environment UBasic \cite{UBasic} and an
arbitrary precision library MAPM for C/C++ \cite{Mapm}. These tools have been
used to ascertain the energy dependence shown in Fig. 2. One can see a really
good agreement with the Levinson theorem, but to achieve this, the phase shift
has been calculated up to very high energies (note that the energy scale in
Fig. 2 is logarithmic, and it involves 20 orders of magnitude!). A good point
is that there is no need for cumbersome high-precision calculations at very
high energies, because this part of the curve can be accurately ascertained
from general considerations. Namely, from Eq. (12), one gets $\delta
(k)=\lim\limits_{r\rightarrow\infty}\delta(r,k)=-\dfrac{1}{2Ck}\int
\limits_{0}^{\infty}V(r)\left\{  1-\cos\left[  2kr+2\delta(r,k)\right]
\right\}  dr$ ($C\equiv\dfrac{\hbar^{2}}{2m}$). Repeatedly integrating by
parts and assuming $k\rightarrow\infty$, the following asymptotic formula can
be obtained%
\begin{equation}
\delta(k)=\dfrac{a_{1}}{k}+\frac{a_{3}}{k^{3}}+\frac{a_{5}}{k^{5}}+...,\text{
}k\rightarrow\infty,
\end{equation}
where $a_{1}=-\dfrac{\int\limits_{0}^{\infty}V(r)dr}{2C}$ and other
coefficients $a_{3},$ $a_{5},...$ are also directly related to the potential
and its derivatives (see \cite{SelgArXiv1} for more details). Using Eq. (13)
one can extend calculations to arbitrarily high energies, and thus, as a
particular result, explicitly verify the Levinson theorem. Since the validity
of Levinson theorem is beyond any doubts, Fig. 2 actually demonstrates the
success of the described approach to calculating the phase shift.

It might be of interest to compare the exact phase shift with the relevant
quantity obtained from a semi-classical (WKB) calculation. In the framework of
the WKB approximation, the phase shift reads \cite{Landau}%
\begin{equation}
\delta(E)=\int\limits_{r_{0}}^{\infty}\left[  \sqrt{\frac{E-V(r)}{C}-\frac
{1}{4r^{2}}}-k\right]  dr+\frac{\pi}{4}-kr_{0},
\end{equation}
where $k=\sqrt{\frac{E}{C}}$ and $r_{0}$ is the solution of the equation
$\dfrac{E-V(r)}{C}=\dfrac{1}{4r^{2}}.$ In Fig. 3 one can see a part of the
exact phase curve in comparison with the WKB phase shift, calculated according
to Eq. (14). As expected, semi-classical model becomes more and more reliable
with increasing energy, but it fails to adequately describe the energy range
$E\lesssim V(0)$ where the exact phase can be ascertained very easily.

\section{Preparatory steps to apply Krein method}

How could one get any practical benefit from full knowledge of the phase shift
and other spectral characteristics related to the reference potential? Trying
to answer this question, let us introduce another important spectral
characteristic, Jost function $F(k)\equiv\left\vert F(k)\right\vert
\exp\left[  -i\delta(k)\right]  $, which fixes relationship between regular
and physical solutions of the Schr\"{o}dinger equation: $\Psi(r,k)=\dfrac
{k}{F(k)}\varphi(r,k).$ The modulus of Jost function reads \cite{Chadan}%
\begin{equation}
\left\vert F(E)\right\vert =\prod\limits_{n=0}^{N}(1-E_{n}/E)\exp\left[
-\frac{1}{\pi}P\int\limits_{0}^{\infty}\frac{\delta(E^{\prime})dE}{E^{\prime
}-E}\right]  ,\text{ }E\in(0,\infty).
\end{equation}
Here $E_{n}$ are the discrete energy eigenvalues and symbol $P$ denotes the
principal value of the integral. Next, we can fix spectral density
\begin{equation}
\dfrac{d\rho(E)}{dE}=\left\{\genfrac{}{}{0pt}{}{\pi^{-1}\sqrt{E}\left\vert F(E)\right\vert 
^{-2},\text{{}}E\geq0,}{\sum\limits_{n}C_{n}\delta(E-E_{n}),\text{ \ }E<0.}\right.
\end{equation}
which is closely related to the real solution schemes of the inverse problem.
An interesting option, among others, is to calculate an auxiliary potential
$V_{0}(r)$ with no bound states, whose spectral density for $E\geq0$ is
exactly the same as that of the reference potential ($\left\vert
F(E)\right\vert $ remains the same). We first show how this problem could be
solved and then, in the next section, discuss of the usefulness of the
approach for practical purposes. To calculate $V_{0}(r)$, one can apply Krein
method \cite{Krein1,Krein2}, based on the following integral equation:%
\begin{equation}
\Gamma_{2r}(r^{\prime})+H(r^{\prime})+\int\limits_{0}^{2r}\Gamma
_{2r}(s)H(s-r^{\prime})ds=0
\end{equation}
whose kernel reads
\begin{equation}
H(r)\equiv\pi^{-1}\int\limits_{0}^{\infty}\left[  \dfrac{1}{\left\vert
F(k)\right\vert ^{2}}-1\right]  \cos(kr)dk.
\end{equation}
The desired potential with no bound states then becomes \cite{Chadan}%
\begin{equation}
V_{0}(r)=4C\left\{  \left[  G(x)\right]  ^{2}-\frac{dG(x)}{dx}\right\}
,\text{ }x\equiv2r,\text{ }G(x)\equiv\Gamma_{2r}(2r).
\end{equation}

Since Eq. (16) is a Fredholm integral equation of second kind, its solution
can be given as a Neumann series \cite{Neumann}%

\begin{gather}
G(x)=-H(x)+\int\limits_{0}^{x}H(x-x_{1})H(x_{1})dx_{1}-\int\limits_{0}^{x}%
\int\limits_{0}^{x}H(x-x_{2})H(x_{2}-x_{1})H(x_{1})dx_{2}dx_{1}+...\nonumber\\
+(-1)^{n}\int\limits_{0}^{x}\int\limits_{0}^{x}...\int\limits_{0}^{x}%
H(x-x_{n})H(x_{n}-x_{n-1})...H(x_{2}-x_{1})H(x_{1})dx_{n}dx_{n-1}...dx_{1}+...
\end{gather}
This general formula can be easily proved, if one uses a suitable quadrature
rule to discreticize Eq. (17), then finds the last element of the solution
vector (the only one which is actually needed) of the corresponding matrix
equation, and finally goes to the limit $h\rightarrow0$, where $h$ is the step
of integration. For relatively small distances the discrete analogue of Eq.
(17) can be solved by Gaussian elimination \cite{SelgArXiv2}, but with
increasing $x$ the number of equations rapidly becomes too large for this kind
of approach. In this situation, one is guided to multidimensional Monte Carlo
techniques using, for example, Metropolis algorithm \cite{Metropolis}. This
interesting possibility, however, is a subject for a forthcoming study and
will be not further discussed here.

Calculation of Jost function and the related function $g(k)\equiv\dfrac
{1}{\left\vert F(k)\right\vert ^{2}}-1$ is described elsewhere
\cite{SelgArXiv1,SelgArXiv2}. As can be seen in Fig. 4, $g(k)=-1$ (with high
accuracy) in a wide range below (and not too close) to the characteristic wave
number $k_{0}=\sqrt{2mV(0)}/\hbar,$ and accurate analytic approximations can
be obtained for the remaining part of the curve. Again, as for the phase
shift, a simple asymptotic formula%
\begin{equation}
g(k)=\dfrac{b_{1}}{k^{2}}+\frac{b_{2}}{k^{4}}+\frac{b_{3}}{k^{6}},\text{
}k\geq k_{a}%
\end{equation}
can be derived, where the leading coefficient $b_{1}=-\dfrac{V(0)}{2C},$ and
the value $k_{a}=75000$ \AA $^{-1}$ has been fixed for the model system
(Xe$_{2}$) studied. Since $H(r)=H(-r)$ is just the Fourier cosine transform of
the characteristic function $g(k),$ calculation of this kernel in the
short-distance range can be performed quite easily and accurately. As shown in
Fig. 5, $H(r)$ is a rapidly oscillating function with decaying amplitude. At
longer distances, however, when the amplitude of $H(r)$ becomes very small,
even minor inaccuracies in calculating $g(k)$ result in large relative errors
of its Fourier transform. Fortunately, a correct asymptotic formula for $H(r)$
can be fixed, namely%
\begin{equation}
H(r)\approx a\exp(-br)\cos(\bar{k}r+\alpha),\text{ }r\rightarrow\infty,
\end{equation}
where $a,b,\bar{k}$ and $\alpha$ are some constants to be specified. To prove
Eq. (22), let us assume that $k\rightarrow\infty,$ so that according to Eq.
(21) one can approximate $g(k)=\dfrac{b_{1}}{k^{2}}.$ We can perform the
inverse Fourier transform $g(k)=2$ $\int\limits_{0}^{\infty}H(r)\cos(kr)dr,$
and divide the domain into two intervals, (0, $r_{a}$] and [$r_{a},\infty)$,
where $r_{a}$ is an arbitrary (but sufficiently large) quantity, such that
$\sin(kr_{a})=0$ and $\cos(kr_{a})=1$ (these conditions can be always set,
because $k$ is arbitrary). Then, taking account of the general formula
$H^{\prime}(0)=-\dfrac{b_{1}}{2}=\dfrac{V(0)}{4C}$\cite{SelgArXiv2},
integrating by parts in the first interval, and using Eq. (22) for the second
interval, one comes to a relation $H^{\prime}(r_{a})+ak^{2}\int\limits_{r_{a}%
}^{\infty}\exp(-br)\cos(\bar{k}r+\alpha)\cos(kr)dr=0.$ It can be explicitly
checked that this relation holds as $k\rightarrow\infty,$ which completes the proof.

Since $\left\vert F(0)\right\vert $ is an immensely large quantity
\cite{SelgArXiv1}, $g(0)=2$ $\int\limits_{0}^{\infty}H(r)dr=-1$ (with very
high precision). Let $x$ be an arbitrary distance, such that Eq. (22) can be
used to calculate $H(x).$ Then%
\begin{equation}
\int\limits_{0}^{x}H(r)dr+1/2=-\int\limits_{x}^{\infty}H(r)dr=\frac
{a\exp(-bx)}{\sqrt{b^{2}+\bar{k}^{2}}}\cos(\bar{k}x+\alpha+\beta),
\end{equation}
where $\cos\alpha=\dfrac{b}{\sqrt{b^{2}+\bar{k}^{2}}}$ and $\sin\alpha
=\dfrac{\bar{k}}{\sqrt{b^{2}+\bar{k}^{2}}}.$ Thereafter, accurately evaluating
the integral $\int\limits_{0}^{x}H(r)dr,$ the desired parameters can be easily
fixed by direct least-squares fit of Eq. (23). To be more specific, Simpson
rule with integration step $h=10^{-7}$ \AA \ was used for $x=0.038$ \AA , and
as a result, the following set of parameters has been ascertained: $a=2.7125$
1/\AA $,$ $b=70.04656$ 1/\AA $,$ $\alpha=-3.67004,$ and most importantly,
$\bar{k}=19165.07$ 1/\AA \ $\approx0.99866$ $k_{0}.$ Fig. 6 demonstrates how
nicely Eq. (22) already works at $r\gtrsim0.03$ \AA . Thus, for this region
there is no need for troublesome Fourier transform to calculate the kernel of
the Krein equation, which makes Eq. (20) much more useful for practical purposes.

\section{Improving the initial potential}

Having ascertained all spectral characteristics of the reference potential and
described a possible solution scheme of the Krein equation, one cannot ignore
the principle question: what is the practical point of all this? Can these
data be used to calculate an improved potential for the system? First, we have
to point out that such an improved (in some sense) potential $V_{1}(r)$ can be
ascertained without auxiliary calculations described in Section III. Most
easily this can be demonstrated in the frame of Gelfand-Levitan method
\cite{GL}, which is based on the integral equation%

\begin{equation}
K(r,r^{\prime})+G(r,r^{\prime})+\int\limits_{0}^{r}K(r,s)G(s,r^{\prime})ds=0,
\end{equation}
whose kernel can be directly linked with the reference potential $V(r)$
\begin{equation}
G(r,r^{\prime})=\int\limits_{-\infty}^{\infty}\left[  d\rho^{(1)}%
(E)-d\rho^{(0)}(E)\right]  \varphi(r,E)\varphi(r^{\prime},E).
\end{equation}
Here
\begin{equation}
\dfrac{d\rho^{(1)}(E)-d\rho^{(0)}(E)}{dE}=\left\{\genfrac{}{}{0pt}{}{\pi^{-1}\sqrt{E}\left[  \left\vert F^{(1)}(E)\right\vert
^{-2}-\left\vert F(E)\right\vert ^{-2}\right]  ,\text{ \ \ \ \ \ \ \ \ }%
E\geq0,}{\sum\limits_{k}C_{k}^{(1)}\delta(E-E_{k}^{(1)})-\sum\limits_{m}%
C_{m}^{(0)}\delta(E-E_{m}^{(0)}),\text{ \ }E<0,}\right.
\end{equation}
the regular solution $\varphi(r,E)$ as well as the quantities $\rho^{(0)}(E),$
$\left\vert F^{(0)}(E)\right\vert ,$ $E_{m}^{(0)}$ and $C_{m}^{(0)},$
($m=1,2,...,n$) correspond to the reference potential $V(r),$ while
$d\rho^{(1)}(E),$ $\left\vert F^{(1)}(E)\right\vert ,$ $E_{k}^{(1)}$ and
$C_{k}^{(1)}$ ($k=1,2,...,l$) are related to $V_{1}(r)$. Suppose we are
looking for a potential $V_{1}(r)$ whose spectral density at $E\geq0$ is
exactly the same as that for the reference potential $V(r)$, i.e., $\left\vert
F^{(1)}(E)\right\vert =\left\vert F(E)\right\vert ,$ but the discrete energy
spectrum and the related norming constants are different. In this case the
kernel is separable:%
\begin{equation}
G(r,r^{\prime})=\sum\limits_{k}C_{k}^{(1)}\varphi\left(  r,E_{k}^{(1)}\right)
\varphi\left(  r^{\prime},E_{k}^{(1)}\right)  -\sum\limits_{m}C_{m}%
^{(0)}\varphi\left(  r,E_{m}^{(0)}\right)  \varphi\left(  r^{\prime}%
,E_{m}^{(0)}\right)  ,
\end{equation}
and Eq. (24) can be easily solved. Thus, in principle, one can directly
calculate the difference of the two potentials in question (see, e.g.,
\cite{Chadan} for more details)%
\begin{equation}
V_{1}(r)-V(r)=2C\frac{d}{dr}K(r,r)=-2C\left\{  \ln\left[  \det S^{(n+l)}%
(r)\right]  \right\}  ^{\prime\prime},
\end{equation}
where the $(n+l)\times(n+l)$ matrix $S^{(n+l)}(r)\equiv I+\int\limits_{0}%
^{r}R^{(n+l)}(s)ds,$ $I$ is the corresponding unit matrix, and the elements of
another matrix $R^{(n+l)}(s)$ read%
\begin{equation}
R_{jk}^{(n+l)}(s)=C_{j}\varphi\left(  s,E_{j}\right)  \varphi\left(
s,E_{k}\right)  ,\text{ }(j,k=1,2,...,n+l).
\end{equation}
The quantities $C_{j}$ and $E_{j}$ are defined as follows: $C_{j}\equiv
C_{j}^{(1)}$, and $E_{j}\equiv E_{j}^{(1)}$, if $1\leq j\leq l;$ $C_{j}%
\equiv-C_{j-l}^{(0)}$, and $E_{j}\equiv E_{j-l}^{(0)}$, if $\ l<j\leq n+l.$ If
the parameters $E_{j}^{(1)}$ and $C_{j}^{(1)}$ are taken equal to the actually
observed values of these quantities, the potential $V_{1}(r)$ can be
considered more realistic than the reference potential $V(r).$

According to Eqs. (28) and (29), one only has to accurately evaluate the
regular solutions $\varphi\left(  s,E_{j}\right)  $ $(j=1,2,...,n+l),$ in
order to ascertain the improved potential $V_{1}(r),$ which seems to
essentially reduce the computation time. This imagination, however, is
deceptive, because full knowledge (from 0 to $\infty$) of $n+l$ regular
solutions is in fact equivalent to calculating $n+l$ auxiliary potentials for
the system. Indeed, let us recall that once the ground state wave function
$\Psi_{0}(r)$ (or any other bound state wave function) is known, the potential
$U(r)$ is immediately known as well (up to a constant) \cite{Cooper}. This
general rule directly results from the Schr\"{o}dinger equation, if one takes
the ground state energy as the zero point of the energy scale: $U(r)=C$
$\Psi_{0}^{\prime\prime}(r)/\Psi_{0}(r).$ In the present case we have to take
into consideration that the solutions $\varphi(r,E_{j}^{(1)})$ are not the
real eigenfunctions, if $E_{j}^{(1)}\neq E_{j}^{(0)}.$ Nevertheless, since the
norming constants $C_{j}^{(1)}$ are expected to be known, the regular
eigenfunctions $\varphi_{1}(r,E_{j}^{(1)})$ related to the potential
$V_{1}(r)$ can be easily ascertained using the well-known relations of the
quantum inverse theory \cite{Chadan}. For example, if there was only one bound
state with energy $E_{1}^{(1)}$and norming constant $C_{1}^{(1)},$ then
$\varphi_{1}(r,E_{1}^{(1)})=\varphi(r,E_{1}^{(1)})\left[  1+C_{1}^{(1)}%
\int\limits_{0}^{r}\varphi^{2}(s,E_{1}^{(1)})ds\right]  ^{-1}.$ Thus,
seemingly simple Eq. (28) "hiddenly" contains a procedure of removing all
previous bound states (equivalent to calculating regular eigenfunctions
$\varphi(r,E_{j}^{(0)}),$ $j=1,2,...,n$ ), and introducing new bound states
(equivalent to calculating regular solutions $\varphi(r,E_{j}^{(0)}),$
$j=1,2,...,l$ ) related to the improved potential $V_{1}(r).$

The explicit procedure equivalent to Eq. (28) can be performed as follows
\cite{AM, LP}. Starting from the potential $V_{n}(r)\equiv V(r),$ one first
removes its zeroth level $E_{0}$ and calculates a new potential, isospectral
with the initial one (apart from $E_{0}$)
\begin{equation}
V_{n-1}(r)=V_{n}(r)+2C\left\{  \frac{2\Psi_{0}^{(n)}(r)\left(  \Psi_{0}%
^{(n)}(r)\right)  ^{\prime}}{\int\limits_{r}^{\infty}\left(  \Psi_{0}%
^{(n)}(s)\right)  ^{2}ds}+\left[  \frac{\left(  \Psi_{0}^{(n)}(r)\right)
^{2}}{\int\limits_{r}^{\infty}\left(  \Psi_{0}^{(n)}(s)\right)  ^{2}%
ds}\right]  ^{2}\right\}  .\text{ }%
\end{equation}
Here $\Psi_{0}^{(n)}(r)$ is the eigenfunction of the zeroth level, not
necessarily normalized (note that the norming constant is absent here). The
new set of wave functions for $V_{n-1}(r)$ can also be expressed in terms of
the previous wave functions for $V_{n}(r)$ \cite{LP}. The procedure can be
continued, i.e., one can remove, one by one, all bound states along with
calculating new auxiliary potentials at each step, until he finally comes to
the potential $V_{0}(r)$ with no bound states. Thereafter, one can introduce
new bound states $E_{j}^{(1)}$ with norming constants $C_{j}^{(1)}$in exactly
the same way as described above, i.e.,
\begin{equation}
V_{1}(r)-V_{0}(r)=-2C\left\{  \ln\left[  \det S^{(l)}(r)\right]  \right\}
^{\prime\prime},
\end{equation}
where the $l\times l$ matrix $S^{(l)}(r)\equiv I+\int\limits_{0}^{r}%
R^{(l)}(s)ds,$ and%
\begin{equation}
R_{jk}^{(l)}(s)=C_{j}^{(1)}\varphi(s,E_{j}^{(1)})\varphi(s,E_{k}^{(1)}),\text{
}(j,k=1,2,...,l).
\end{equation}

Alternatively, one can solve Krein equation, as described in Section III, and
then use Eq. (19) to get the auxiliary potential $V_{0}(r)$ with no bound
states. As mentioned, all three approaches to calculating the improved
potential $V_{1}(r)$ are equivalent, in presumption that the modulus of Jost
function remains unchanged, which means that \cite{Chadan}%
\begin{equation}
F_{1}(k)=\prod\limits_{j=1}^{l}\left(  \frac{k-ik_{j}^{(1)}}{k-ik_{j}^{(1)}%
}\right)  \cdot F_{0}(k),
\end{equation}
and
\begin{equation}
F(k)=\prod\limits_{j=1}^{n}\left(  \frac{k-ik_{j}^{(0)}}{k-ik_{j}^{(0)}%
}\right)  \cdot F_{0}(k),
\end{equation}
where $F_{1}(k),$ $F(k)$ and $F_{0}(k)$ are the Jost functions for the
potentials $V_{1}(r),$ $V(r)$ and $V_{0}(r)$, respectively, $k_{j}^{(1)}%
=\sqrt{2mE_{j}^{(1)}}/\hbar$ and $k_{j}^{(0)}=\sqrt{2mE_{j}^{(0)}}/\hbar.$

An alternative to applying Eq. (31) is to introduce new bound states
step-by-step, starting from the auxiliary potential $V_{0}(r).$ This
well-known procedure is based on the formula \cite{Chadan} (see also
\cite{SelgArXiv2})
\begin{equation}
V_{k}(r)=V_{k-1}(r)-2C\left\{  \ln\left[  1+C_{k}^{(1)}\int\limits_{0}%
^{r}\varphi_{k-1}^{2}\left(  s,E_{k}^{(1)}\right)  ds\right]  \right\}
^{\prime\prime},\text{ (}k=1,2,...,l\text{)}%
\end{equation}
where the regular solution $\varphi_{k-1}(s,E_{k}^{(1)})$ is not the real
eigenfunction, since it is related to the potential $V_{k-1}(r).$ Naturally,
the norming constants $C_{j}^{(1)}$ in Eqs. (29), (32) and (35) are different
(although the same notation is used), because they are related to different potentials.

Currently, one can only guess which of the described approaches is
computationally most effective. However, one cannot argue that solution of
Krein equation is an interesting problem on its own, and the scheme outlined
in Section III is realizable. At least, it seems reasonable to first calculate
the auxiliary potential $V_{0}(r)$ with no bound states, and then try to
improve the initial potential. Introducing "real" bound states is not the only
option in this context. For example, one can construct another potential
$\bar{V}_{0}(r)$ with no bound states, a Bargmann potential whose Jost
function reads%
\begin{equation}
\bar{F}_{0}(k)=\prod\limits_{j=1}^{m}\left(  \frac{k+ia_{j}}{k+ib_{j}}\right)
\cdot F_{0}(k),
\end{equation}
where the quantities $a_{j}$ $\neq b_{j}$ are real and positive. For
simplicity, let us assume that $m=1$ ($a_{1}\equiv a$ and $b_{1}\equiv b$).
Then \cite{Chadan}%
\begin{equation}
\bar{V}_{0}(r)-V_{0}(r)=-2C\left\{  \ln\frac{W\left[  f_{1}(r,ia),\varphi
_{1}(r,ib)\right]  }{b^{2}-a^{2}}\right\}  ^{\prime\prime},
\end{equation}
where the symbol $W$ denotes Wronskian determinant, and $f_{1}(ia,r)$ is the
Jost solution of the Schr\"{o}dinger equation ($f_{1}(r,ia)\rightarrow
\exp(-ar)$ as $r\rightarrow\infty$). Both $f_{1}(r,ia)$ and the regular
solution $\varphi_{1}(r,ib)$ are related to the potential $V_{0}(r).$ Thus,
suitably choosing the parameters $a_{j}$ and $b_{j}$, one can, in principle,
take a more adequate account of the scattering data or of the known behavior
of the potential near the origin. Suppose, for example, that the quantity
$\bar{V}_{0}(0)$ is known. Then, using the formula%
\begin{equation}
\frac{d}{dr}\left[  \frac{W\left[  f_{1}(r,ia),\varphi_{1}(r,ib)\right]
}{b^{2}-a^{2}}\right]  =-f_{1}(r,ia),\varphi_{1}(r,ib),
\end{equation}
which is valid for any two solutions of the Schr\"{o}dinger equation for the
same potential, taking account that $\left\vert F_{0}(k)\right\vert
^{2}=1+\dfrac{V(0)}{2Ck^{2}}+0(k^{-4})$ \cite{SelgArXiv1}, and consequently,
$\left\vert \bar{F}_{0}(k)\right\vert ^{2}=1+\dfrac{\bar{V}(0)}{2Ck^{2}%
}+0(k^{-4}),$ one comes to a more transparent form of Eq. (37):%
\begin{equation}
\frac{\bar{V}_{0}(r)-V_{0}(r)}{\bar{V}_{0}(0)-V_{0}(0)}=\frac{d}{dr}\left[
\frac{f_{1}(r,ia),\varphi_{1}(r,ib)}{W\left[  f_{1}(r,ia),\varphi
_{1}(r,ib)\right]  }\right]  .
\end{equation}
One can easily check that the right side of Eq. (39) indeed approaches unity
as $r\rightarrow0,$ because $\varphi_{1}(0,ib)=0,$ but $f_{1}(0,ia)\neq0$
(otherwise $E_{a}=-Ca^{2}$ would be a bound state).

Using the modified auxiliary potential $\bar{V}_{0}(r),$ one can construct
another improved potential for the system, introducing new bound states all at
once, according to Eq. (31) or step-by-step, according to Eq. (35).

\section{Conclusion}

Starting from a known reference potential, one can calculate important
spectral characteristics of the quantum system, which are almost unattainable
in a real experiment. Then it is possible, in principle, to construct another,
in some sense more realistic potential for the system. This kind of approach
has been applied to diatomic xenon molecule in its ground electronic state.
The phase shift for the full range of scattering states has been calculated,
and excellent agreement with Levinson theorem explicitly demonstrated. All
this important spectral information, along with full discrete energy spectrum,
can be stored into Jost function, and then reused for constructing a new
(possibly improved) potential.

Choice of a proper reference potential, applicable in the whole physical
domain, is a difficult and rather speculative task in any particular case. A
variety of semiclassical schemes have been devised to deduce the potential
directly from the available experimental data, and these concepts are
constantly improved and developed (see, e.g., \cite{LeRoy1, LeRoy2}). Such
methods, although inaccurate from the rigorous quantum mechanical point of
view, have proven useful for practical purposes. Therefore, they might help to
construct a reference potential, which can be used as a reasonable
approximation to the real potential. Only if the reference potential is
realistic enough, there is a chance to construct a Bargmann potential that
would be even more realistic. On the other hand, as demonstrated in this
paper, the computational-technical difficulties can be overcome much more
easily, if one chooses an exactly solvable reference potential (in the sense
specified in Section I). To this end, a multi-component potential of smoothly
joined Morse-type pieces is especially suitable. Full energy dependence of the
phase shift for such a potential can be ascertained analytically, using a
method (described in Section II), which is largely based on a so-called
pseudo-Morse approximation for the small distances (and high energies) region
of the potential. In this paper, only a single pseudo-Morse component has been
used for the range $r\in\lbrack0,X_{1}].$ As demonstrated elsewhere
\cite{Selg1, Selg2}, adding more pseudo-Morse components does not bring along
any serious computational problems, so one can include just as many components
of this type as he considers reasonable.

An important constituent of the overall concept is Krein method for
calculating an auxiliary potential $V_{0}(r)$ with no bound states. For this
purpose one uses the Jost function, which has been ascertained for the
reference potential. General solution of Krein equation, Eq. (20), can be
given as a series of multi-dimensional integrals (Neumann series). Thus, Eq.
(20) represents a formulation of a serious (but very interesting)
computational-technical problem, not solved yet. Fortunately, a simple
asymptotic formula, Eq. (22), for the kernel of Krein equation can be used to
accurately evaluate the multi-dimensional integrands in Eq. (20). As
demonstrated in Section III, Eq. (22) can already be used at rather small
distances, $r\gtrsim0.03$ \AA , instead of performing Fourier cosine transform
of the characteristic function $g(k)$ there.

We discussed several possibilities of improving the initial reference
potential in the rigorous framework of the quantum-mechanical inverse theory.
In particular, explicit calculation of the auxiliary potential $V_{0}(r)$ is
not obligatory for this purpose, if one prefers to calculate regular
eigenfunctions of all bound states for the whole physical domain, and
additionally assumes that $\left\vert F^{(1)}(E)\right\vert =\left\vert
F(E)\right\vert =\left\vert F_{0}(E)\right\vert $ (see Eqs. (33), (34) and the
paragraph between Eqs. (26) and (27)). In this case, one can try to use Eqs.
(28) and (29) to ascertain an improved potential $V_{1}(r)$ whose discrete
energy spectrum would exactly coincide with the "real" one. This method is
extremely difficult to implement, but exactly the same can be said about other
possible approaches described and discussed in Section IV. None of these
schemes has been realized yet, and there is no real ground to give preference
to any of them. In this paper the emphasis has been put on promoting Krein
method to calculating the auxiliary potential with no bound states, which is
an interesting theoretical and computational-technical problem on its own.
Rapid development of parallel computing and Grid technology opens a new
powerful channel for the solution of such complex problems, which suggest
optimism for further research in this direction.

\section*{Acknowledgement}

The research described in this paper has been supported by Grants No 5863 and
5549 from the Estonian Science Foundation.

\pagebreak

\section*{Figure captions}

\begin{enumerate}
\item[Fig. 1.] Three-component model potential for the system (Xe$_{2}$ in
ground electronic state) investigated. All components have the well-known
analytic form of the Morse potential, but the ordinary Morse approximation is
used only in the central range $r\in\left[  X_{1},X_{2}\right]  $ (see the
explanations in Section II). The parameters of the components as well as the
calculated discrete energy levels (24 in total) are given in Ref. 15.

\item[Fig. 2.] Explicit demonstration of the Levinson theorem ($\delta
(0)-\delta(\infty)=n\pi$) for the model system studied. As needed,
$\delta(0)=24\pi$, since the system has 24 bound states. At $E=$ 3.146294 meV,
the phase shift passes a zero, and then remains negative, very slowly
approaching the limit ($\delta(\infty)=0$) as $E\rightarrow\infty.$ The left
side-inset shows the nearly linear energy dependence as $E\rightarrow0$, in
full agreement with general-theoretical concepts. In the right-side inset one
can see that the phase curve has an inflection point near $E=V(0)$.

\item[Fig. 3.] A comparison between the exact phase shift (solid line) and WKB
phase shift (open circles), calculated according to Eq. (14). The difference
between the two phase curves is shown in the inset.

\item[Fig. 4.] Visualization of the characteristic function $g(k)=\left\vert
F(k)\right\vert ^{-2}-1$, which determines the kernel of the Krein equation.
As can be seen, $g(k)=-1$ (with very high precision) if $k\lessapprox
k_{0}=\sqrt{2mV(0)}/\hbar$, then starts to monotonously decrease, and
$g(k)\rightarrow0$ as $k\rightarrow\infty$. In spite of the seeming simplicity
of the energy dependence, the whole curve has to be calculated very accurately
up to high energies, in order to accurately ascertain the kernel $H(r)$. The
high-energy part of the curve ($k\geq k_{a}$) has been calculated according to
Eq. (21). The asymptotic wave number $\bar{k}$ (see Eq. (22) ) is also shown.

\item[Fig. 5.] Demonstration of the behavior of Krein $H$-function (kernel of
Krein equation) given by Eq. (18). The lower graphs start where the upper ones
end. Note that the period of oscillations rapidly achieves the characteristic
value $\dfrac{2\pi}{\bar{k}}$.

\item[Fig. 6.] Demonstration of the validity of the asymptotic formula for
calculating the kernel of Krein equation. Open circles in the figure represent
the result of Fourier cosine transform according to Eq. (18), while the solid
line corresponds to Eq. (22). One can see excellent agreement between the two
principally different approaches to calculating the same quantity.
\end{enumerate}

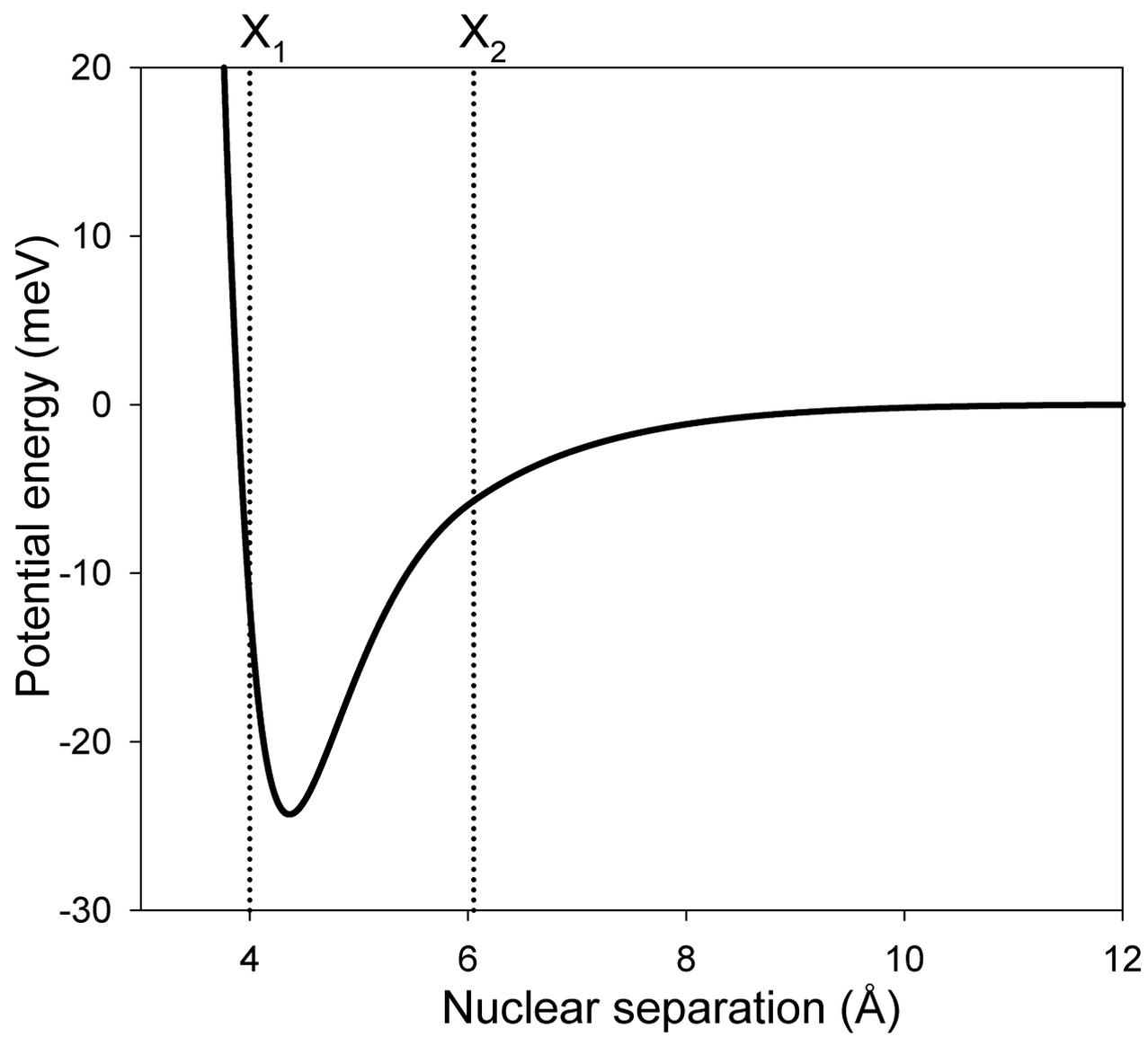
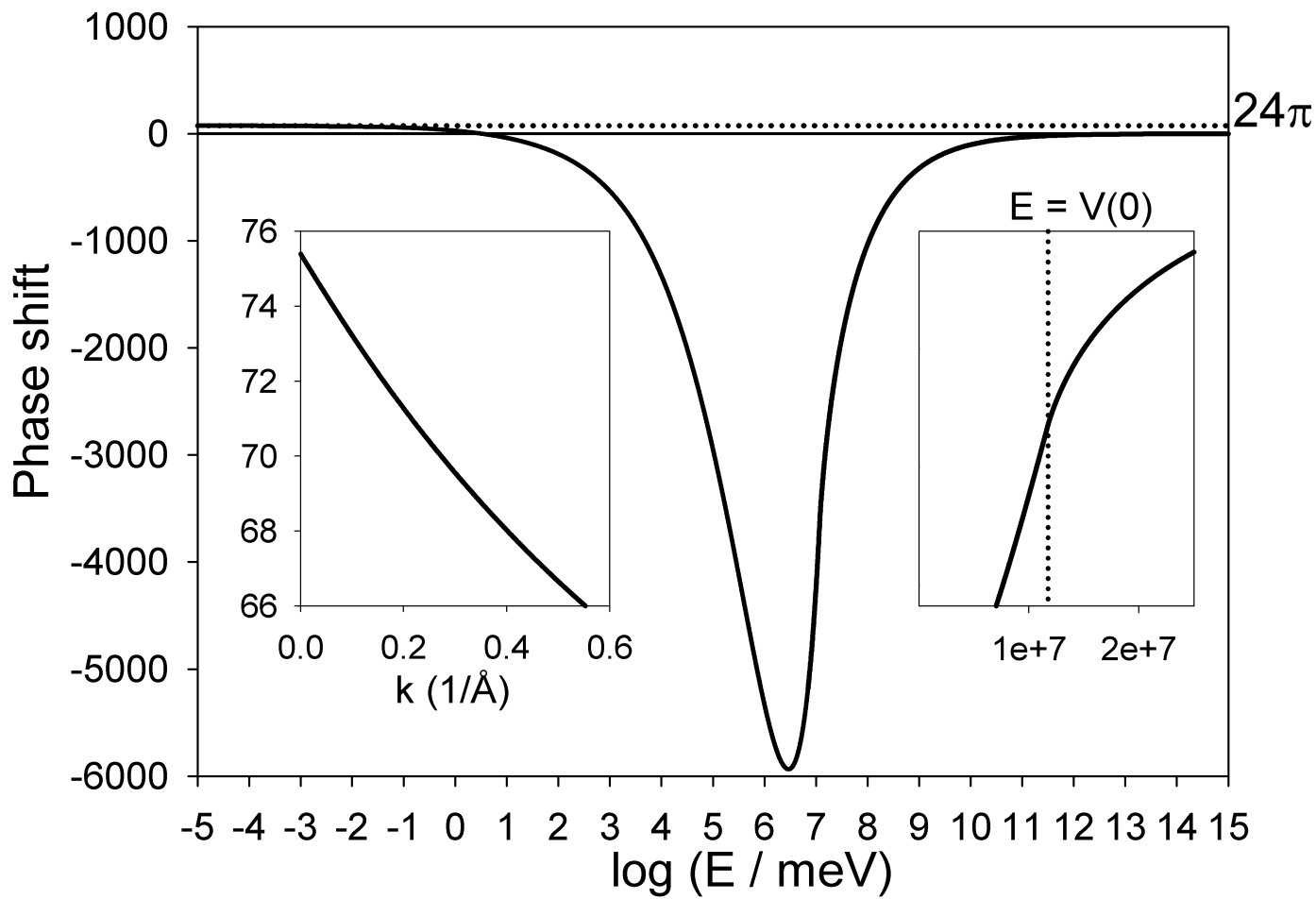
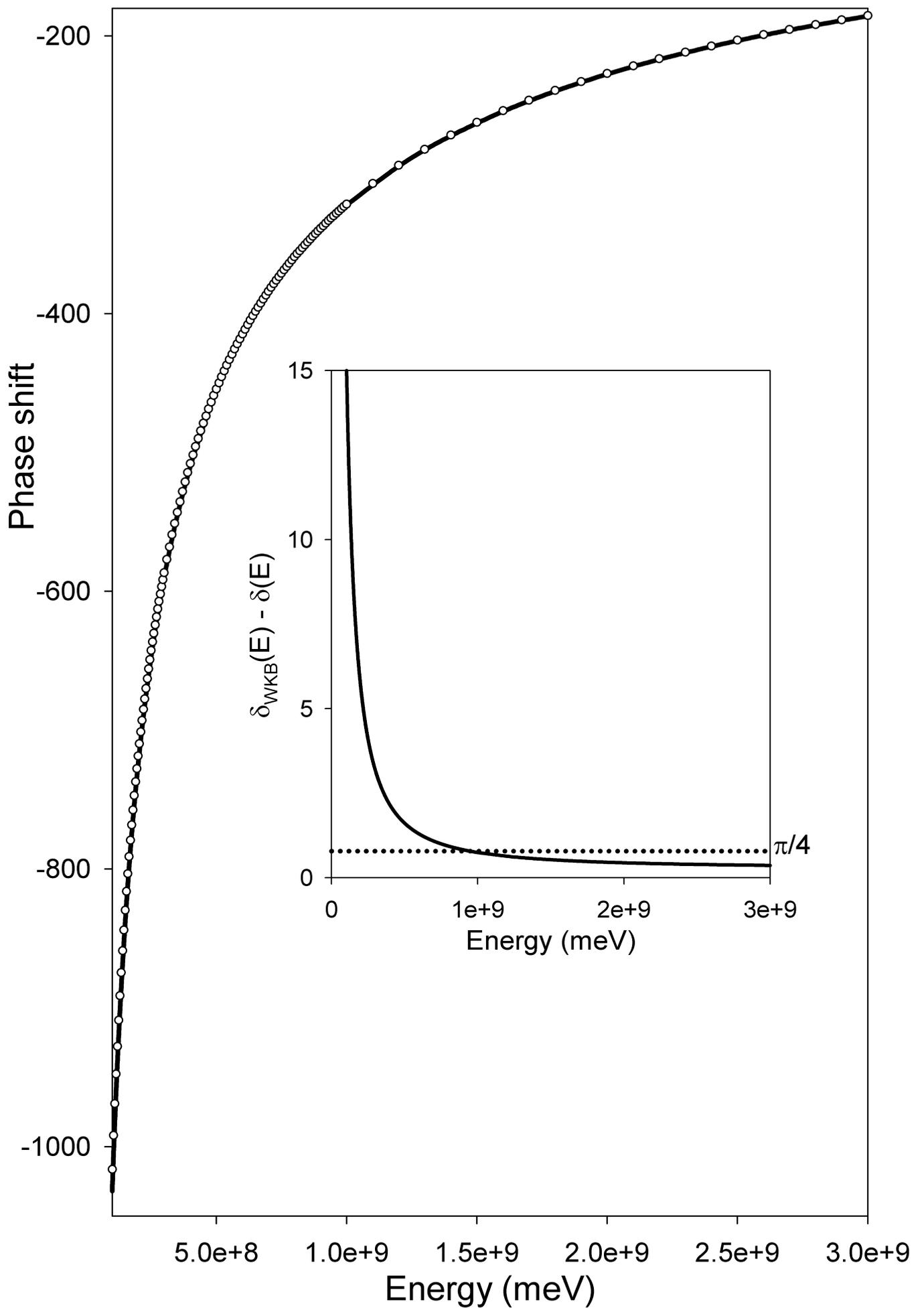
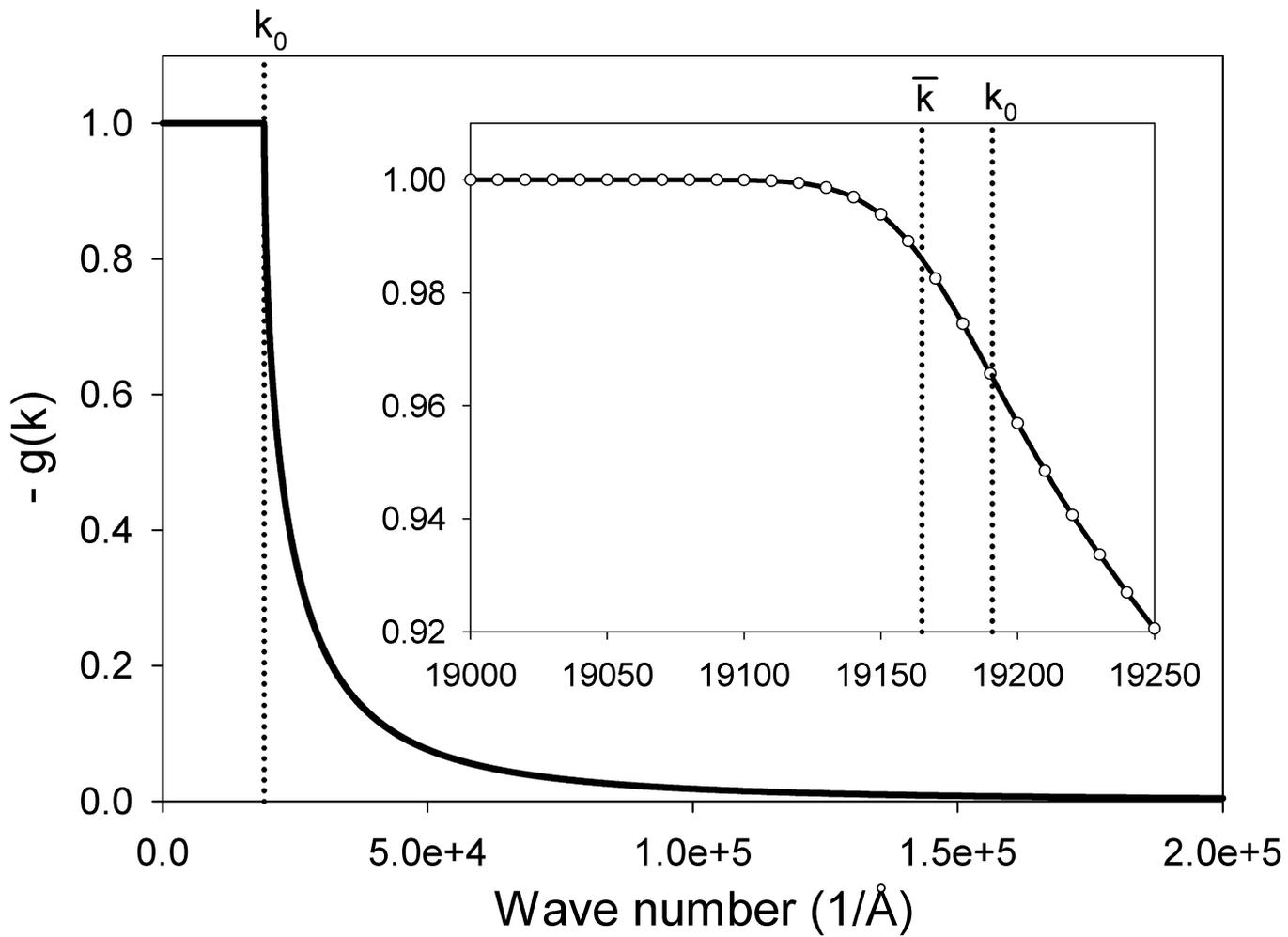
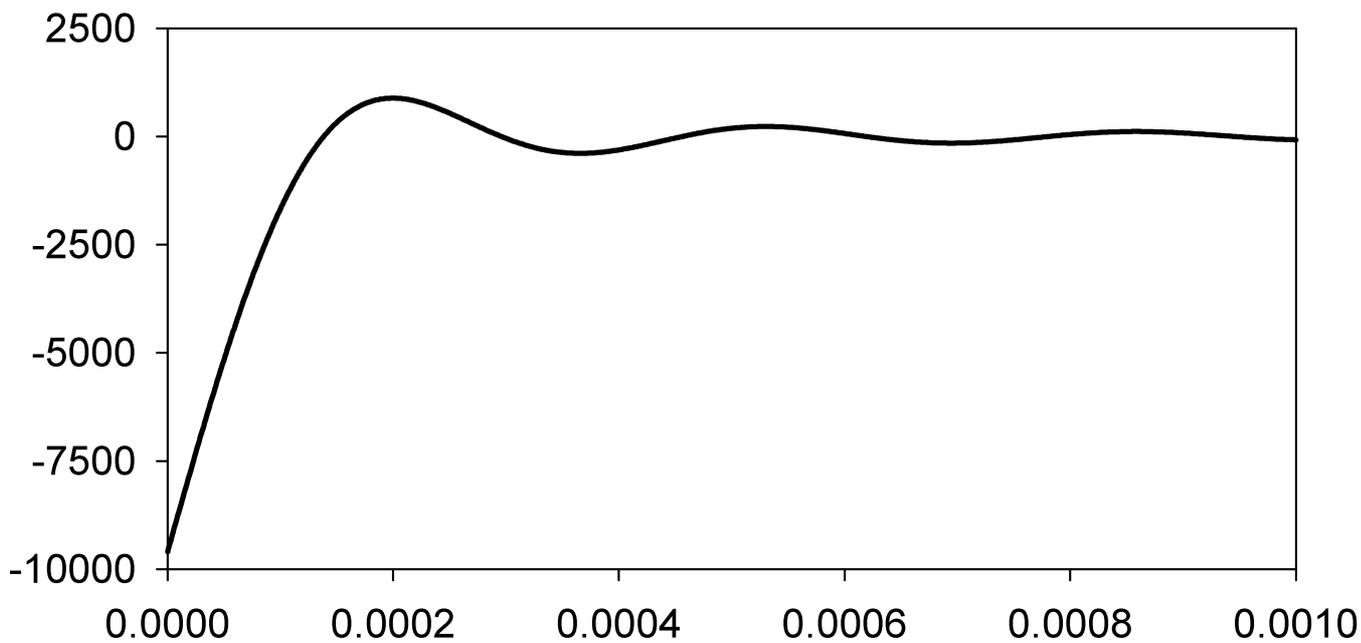
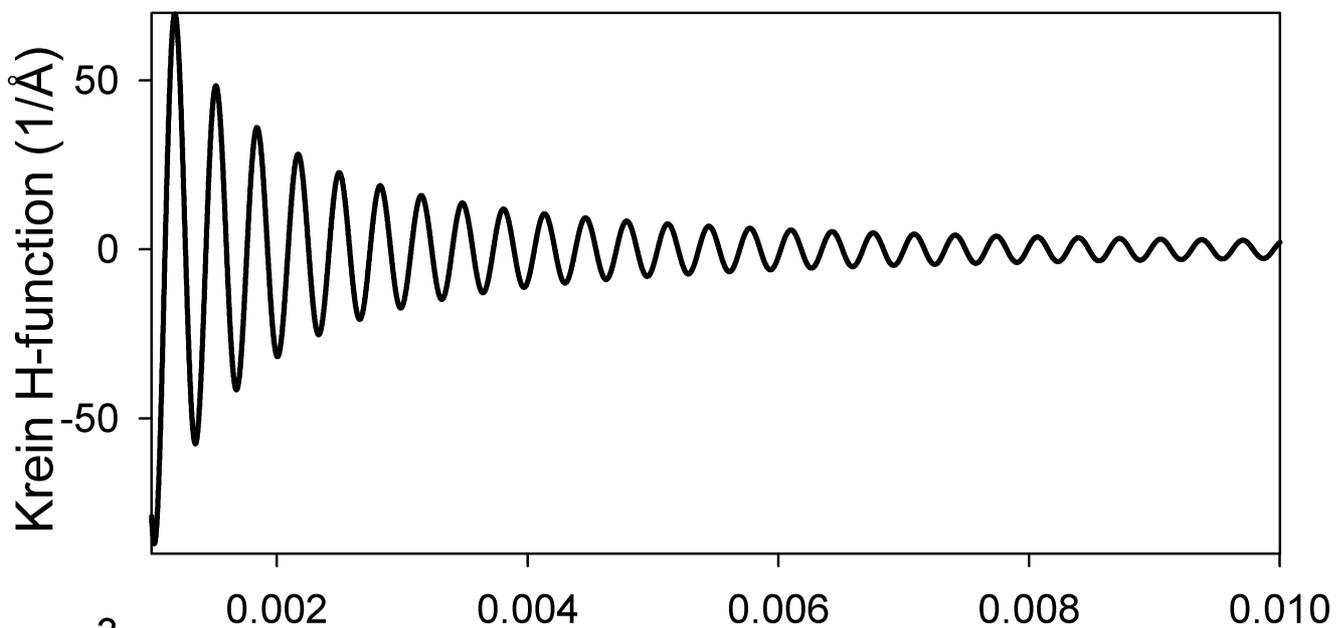
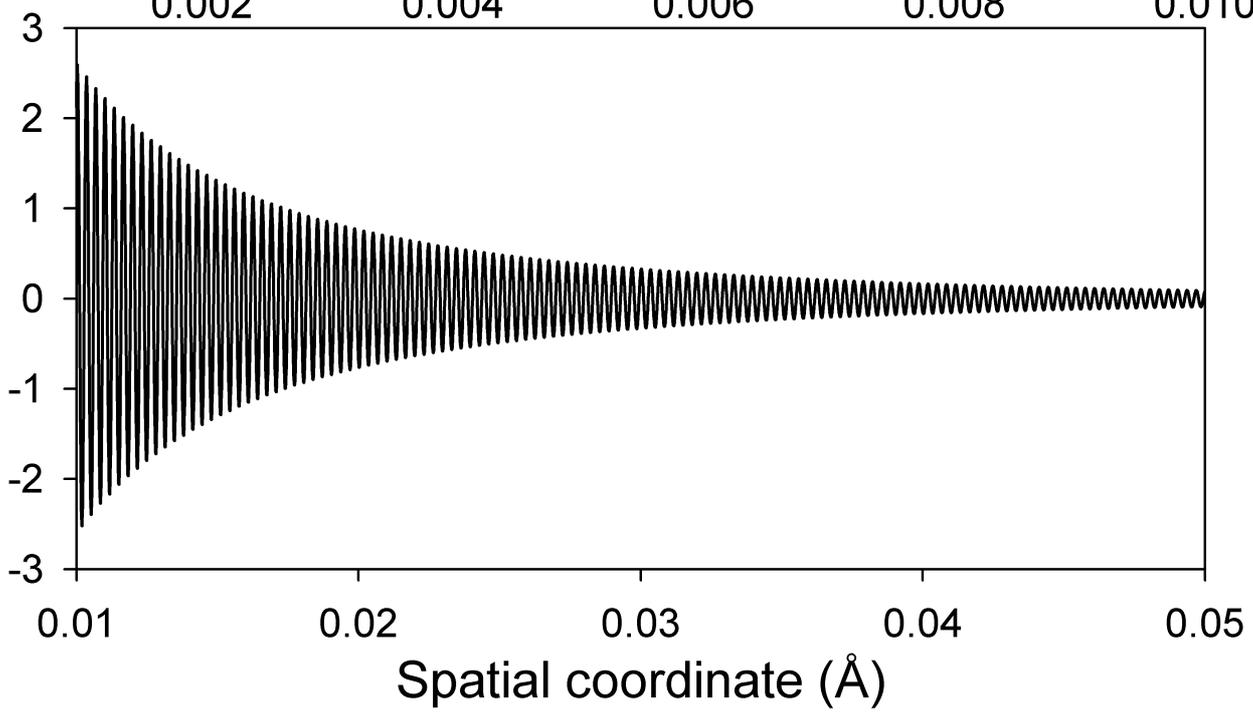
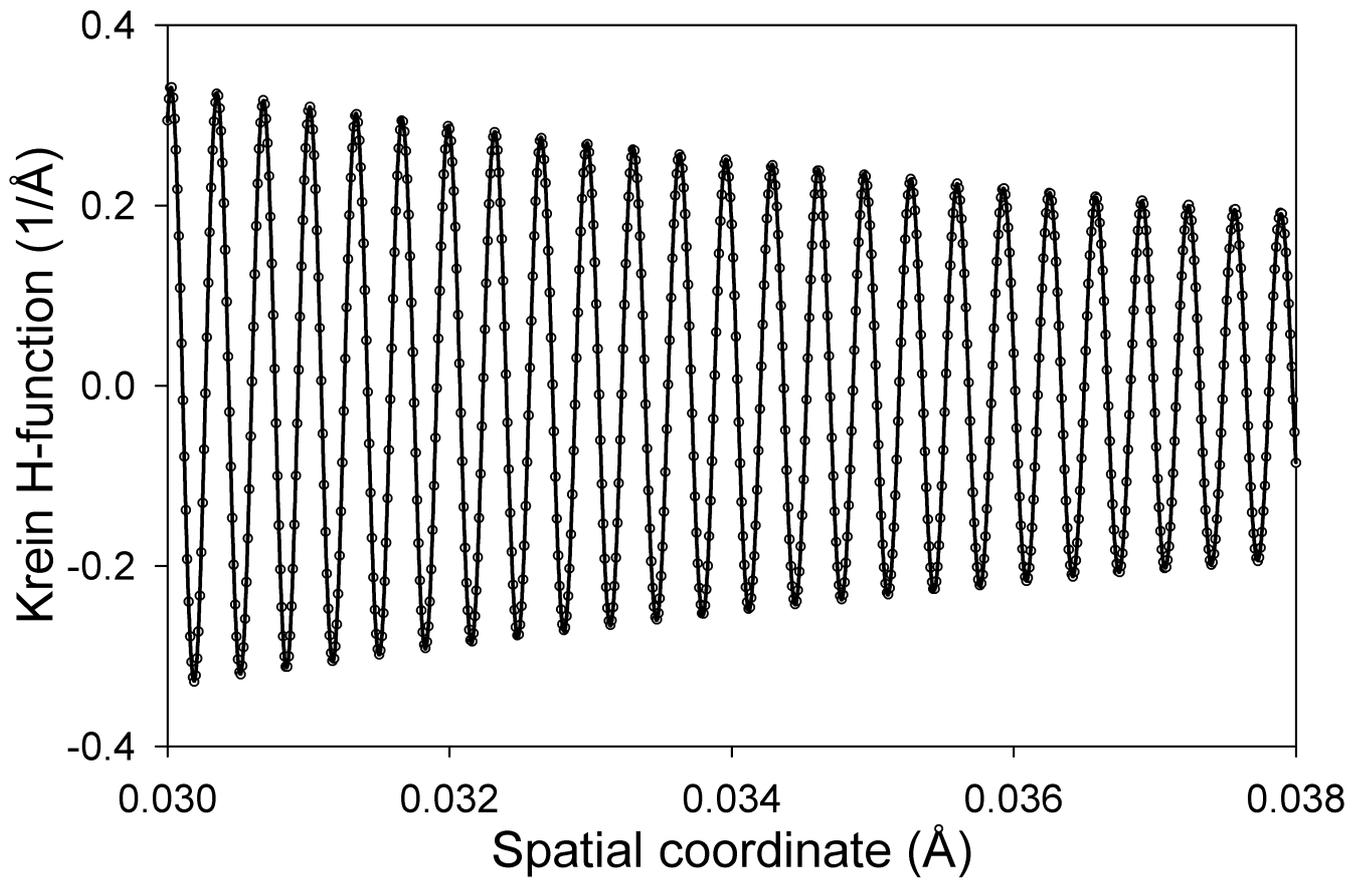
\end{document}